\documentclass[preprint,showpacs,aps]{revtex4}

\usepackage{amsmath}
\usepackage{dcolumn}
\usepackage{verbatim}
\usepackage{mathrsfs}
\usepackage{graphicx}

\begin{document}

\title{Seniority conservation and seniority violation in the $g_{9/2}$ shell}

\author{A.~Escuderos and L.~Zamick}
\affiliation{Department of Physics and Astronomy, Rutgers University,
Piscataway, New Jersey 08854}

\date{\today}

\begin{abstract} 
The $g_{9/2}$ shell of identical particles is the first one for which one can 
have seniority-mixing effects. We consider three interactions: a delta 
interaction that conserves seniority, a quadrupole--quadrupole ($Q\cdot Q$)
interaction that does not, and a third one consisting of two-body matrix 
elements taken from experiment ($^{98}$Cd) that also leads to some seniority 
mixing. We deal with proton holes relative to a $Z=50, N=50$ core. 
One surprising result is that, for a four-particle system with total 
angular momentum $I=4$, there is one state with seniority $v=4$ that is an 
eigenstate of {\it any} two-body interaction---seniority conserving or not. The
other two states are mixtures of $v=2$ and $v=4$ for the seniority-mixing 
interactions. The same thing holds true for $I=6$. Another point of interest is
that, in the single-$j$-shell approximation, the splittings $\Delta E=
E(I_\text{max})-E(I_\text{min})$ are the same for
three and five particles with a seniority conserving interaction (a well known
result), but are equal and opposite for a $Q\cdot Q$ interaction. We also fit 
the spectra with a combination of the delta and $Q\cdot Q$ 
interactions. The $Z=40,N=40$ core plus $g_{9/2}$ neutrons (Zr isotopes) is 
also considered, although it is recognized that the core is deformed.
\end{abstract}


\maketitle

\section{Introduction}

With the advent of nuclei far from stability, there will be more emphasis on 
identical particles in given shells, in which case the concept of seniority 
will be even more important than it has been in the past. See for example the
work of Lisetskiy et al.~\cite{lbhg04}.

Also as part of the revival are works on quasispin and seniority by Rowe and 
Rosensteel~\cite{rr01,rr03} and by the present authors~\cite{ze05}.

There are several well known formulae for seniority selection rules in the
nuclear physics and atomic physics textbooks~\cite{st63,l80,t93}. One of the 
first things we learn 
is that for identical particles in a single $j$ shell, seniority is conserved 
for all shells with $j \le 7/2$, no matter what two-body interaction is used. 
The first shell, then, where seniority violating effects can be seen is the 
$g_{9/2}$, and this is the shell we shall consider here.

As noted in Igal Talmi's review article~\cite{t03}, there have been many 
calculations done in the $g_{9/2}$ region, including calculations with 
seniority-conserving interactions by Gloeckner and collaborators~\cite{gs74,
gl74}, as well as experiment and theory by Oxorn et~al.~\cite{omkw85}. Also to 
be mentioned are Amusa and Lawson~\cite{al82,al83}, and Auerbach and 
Talmi~\cite{at65}.

Our motivation in this work is to see how the effective interaction depends on
what ``closed shell'' is used. 

Some of the well known statements and theorems concerning states of good 
seniority are:
\begin{itemize}
\item[a)] The seniority is roughly the number of identical particles not 
coupled to zero. Hence, for a single nucleon the seniority $v$ is equal to 1. 
For two nucleons in a $J=0$ state we have $v=0$, but for $J=2,4,6$, etc., 
$v=2$. For three nucleons there is one state with seniority $v=1$, which must 
have $J=j$; all other states have seniority $v=3$.

\item[b)] The number of seniority-violating interactions is $\left[ (2j-3)/6
\right]$, where the square brackets mean the largest integer contained therein 
(see Ref.~\cite{gh93}). For $j=7/2$ there are no seniority-violating 
interactions, while for $j=9/2$ there is one.

\item[c)] With seniority-conserving interactions, the spectra of states of the 
same seniority is independent of the particle number.

\item[d)] At midshell we cannot have any mixing of states with seniorities $v$
and $v+2$; one can mix $v$ and $v+4$ states.

\end{itemize}

There are also well known results for more general interactions which do not
necessarily conserve seniority:
\begin{itemize}
\item[a)] For identical particles in a single $j$ shell, the hole spectrum is 
the same as the particle spectrum. This will be relevant to $^{93}$Tc and 
$^{97}$Ag, and for $^{83}$Zr and $^{87}$Zr.

\item[b)] The magnetic moment of a hole is the same as that of a particle. The
quadrupole moment of a hole is equal in magnitude but of the opposite sign to
that of a particle. This leads to the result that at midshell all static 
quadrupole moments vanish in the single-$j$-shell approximation.
\end{itemize}

There have been many calculations in the past in the ``$g_{9/2}$ region'',
although perhaps due to a lack of data, some but not all of the nuclei we 
consider here have been addressed. However, our main motivation is not so much
to do better calculations, but rather to look at the nuclei from a somewhat 
different point of view. As will be seen, there are some surprises even at this
late date concerning $g_{9/2}$ coefficients of fractional parentage. 
Furthermore, while most emphasis has been on seniority-conserving interactions,
there are some simplicities even for seniority-nonconserving interactions, such
as the quadrupole--quadrupole interaction, which we will utilize to determine 
the degree of seniority nonconservation. Also newer data permits us to go 
further away from the valley of stability than was possible at earlier times. 
This enables us to start from different cores and to study the core dependance 
of the effective interaction.

\section{Our calculational method}

We used a program given to us by Bayman to calculate cfp's. However, when there
is more than one state of a given seniority, the Bayman program does not give 
the same cfp's as are recorded in the original Bayman--Lande paper~\cite{bl66}.
Nevertheless, this is not a cause for concern as will be discussed in the next 
section.

What is unusual in our approach is that we do not consider a system of 
identical particles, but rather a system of $(n-1)$ protons and one neutron. 
When we perform the matrix diagonalization, using only isospin-conserving 
interactions, we obtain states with isospin $T_\text{min}=|N-Z|/2$ and also
ones with higher isospins. The latter states are analog states of systems
of identical particles. If we write the wave function as
\begin{equation}
\Psi^I = \sum_{J_P} D^I(J_P v_P, J_N=j) \left[ (j^{n-1})^{J_P} j \right]^I ~,
\end{equation}
then, for the $T_\text{min}+1$ states, the coefficients $D^I$ are coefficients 
of fractional parentage $(j^{n-1} J_P v_P j |\} j^n I)$. We are also interested
in the spectra of $T_\text{min}$ states, but we will save this for another day.

\section{Special behaviours for $I=4^+$ and $6^+$ states of the $g^4_{9/2}$
configuration}

For a system of four identical nucleons in the $g_{9/2}$ shell, the possible 
seniorities are $v=0,2$, and 4, with $v=0$ occurring only for a state of total 
angular momentum $I=0$. There is also a $v=4$ state with $I=0$.

For $I=4$ and 6, we can have three states, one with seniority $v=2$ and two 
with seniority $v=4$. For the two $v=4$ states we have at hand, we can 
construct different sets of $v=4$ states by taking linear combinations of the 
original ones. If the original ones are $(4)_1$ and $(4)_2$, we can form
\begin{eqnarray}
(4)_A & = & a (4)_1 + b (4)_2 ~, \nonumber \\ 
(4)_B & = & -b (4)_1 + a (4)_2 ~,
\end{eqnarray}
with $a^2+b^2=1$. The set $(4)_A$, $(4)_B$ is as valid as the original set.

However, we here note that if we perform a matrix diagonalization with any 
two-body interaction---seniority conserving or not---, one state emerges which 
does not depend on what the interaction is. The other two states are, in 
general, mixtures of $v=2$ and $v=4$ which do depend on the interaction. 
The values of the coefficients of fractional parentage (cfp's) of this unique
state of seniority 4 are shown in Table~\ref{tab:uni-v4}. The states with $J_0 
\ne 4.5$ all have seniority $v=3$. Note that in this special $v=4$ state there 
is no admixture of states with $J_0=j=9/2$, be they $v=1$ or $v=3$. Again, no 
matter what two-body interaction is used, this $I=4$ state remains a unique 
state.

\begin{table}[ht]
\caption{A unique $J=4,v=4$ cfp for $j=9/2$.}\label{tab:uni-v4}
\begin{center}
\begin{tabular*}{.5\textwidth}[t]{@{\extracolsep{\fill}}dd}
\toprule
\multicolumn{1}{c}{$J_0$} & \multicolumn{1}{c}{$(j^3 J_0 j |\} j^4~I=4, v=4)$} 
\\
\colrule
1.5 & 0.1222 \\
2.5 & 0.0548 \\
3.5 & 0.6170 \\
4.5~(v=1) & 0.0000 \\
4.5~(v=3) & 0.0000 \\
5.5 & -0.4043 \\
6.5 & -0.6148 \\
7.5 & -0.1597 \\
8.5 & 0.1853 \\
\botrule
\end{tabular*}
\end{center}
\end{table}

Amusingly, this state {\it does not appear} in the compilation of 
seniority-classified cfp's of Bayman and Lande~\cite{bl66} or de~Shalit and 
Talmi~\cite{st63}.
We should emphasize that, although different, the Bayman--Lande cfp's are 
perfectly correct (as are the ones of de~Shalit and Talmi, whose cfp's are also
different from those of Bayman and Lande~\cite{bl66}). But then, why do they 
not obtain the unique state that we have shown above? Bayman and Lande use 
group theoretical techniques to obtain the cfp's, diagonalizing the following
Casimir operator for $Sp(2j+1)$
\begin{equation}
G(Sp_{2j+1}) = \frac{1}{2j+1} \sum_{\text{odd } k=1}^{2j} (-1)^k 
(2k+1)^{3/2} \left[ U^k U^k \right]^0_0 ~,
\end{equation}
where $U^k_q \equiv \sum_{i=1}^N U^k_q (i)$ and $U$ is the Racah unit tensor
operator
\begin{equation}
\langle \Psi^{j'}_{m'} | U^k_q | \Psi^j_m \rangle = \delta_{j j'}
(k j q m | j' m') ~.
\end{equation}
The two seniority $v=4$ states are degenerate with such an interaction and, 
since there is no seniority mixing, we can have arbitrary linear combinations 
of the $4^+$ states. Only by using an interaction which removes the degeneracy 
and violates seniority, do we learn about the special state in 
Table~\ref{tab:uni-v4}.

\section{The energy splitting $E(I_\text{max})-E(I_\text{min})$ with a $Q\cdot
Q$ interaction}

A well known result for identical particles in a single $j$ shell is that, if 
one uses a seniority-conserving interaction, then the relative spectra of 
states of the same seniority are independant of the number of 
particles~\cite{st63,l80,t93}. Thus, for $n=3$ and $n=5$, the seniority $v=3$ 
states have the same relative spectrum; for $n=2,4$, and 6, the seniority $v=2$
states have the same spectrum. These results hold, in particular, for the delta
interaction used here.

Now the $Q\cdot Q$ interaction does not conserve seniority and the above 
results do not hold. However, we have noticed an interesting result for $n=3$
and $n=5$. Consider the splitting $\Delta E=E(I_\text{max})-E(I_\text{min}),
~v=3$, where for $g_{9/2}$, $I_\text{max}=21/2$ and $I_\text{min}=3/2$. For a
seniority-conserving interaction, $\Delta E(n=5)=\Delta E(n=3)$, whereas for
a $Q \cdot Q$ interaction, $\Delta E(n=5)=-\Delta E(n=3)$. This will be 
discussed quantitatively later.

\section{The two-particle (two-hole) systems}

\begin{figure}[ht]
\includegraphics[scale=.5,clip]{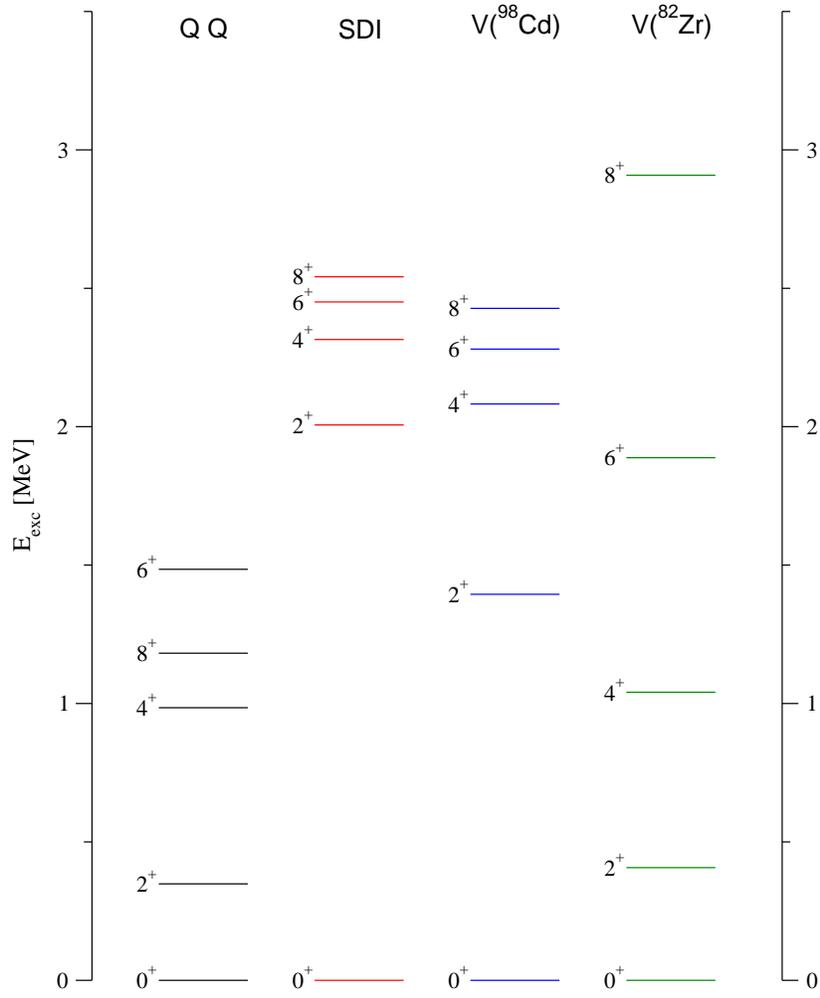}
\caption{Values of the even-$J$ two-body matrix elements (m.e.) for the 
following interactions: quadrupole--quadrupole ($Q\cdot Q$), surface delta 
(SDI), m.e. taken from experimental spectrum of $^{98}$Cd [V($^{98}$Cd)], and 
m.e. taken from experimental spectrum of $^{82}$Zr [V($^{82}$Zr)].}
\label{fig:exp-the}
\end{figure}

In order to perform calculations, we must know the two-body matrix elements 
$E(J)= \langle (g_{9/2}^2)^J |V| (g_{9/2}^2)^J \rangle$. Since in this work we 
consider only two identical nucleons, we need simply the even-$J$ matrix 
elements ($J=0,2,4,6$, and 8). In Table~\ref{tab:3} and Fig.~\ref{fig:exp-the} 
we give four sets of two-body matrix elements. These are: seniority-violating 
quadrupole--quadrupole interaction, seniority-conserving delta interaction, 
matrix elements taken from the two-proton-hole system (relative to a $Z=50,
N=50$ core) $^{98}$Cd, and matrix elements taken from the two-neutron-particle
system (relative to $Z=40,N=40$) $^{82}$Zr.

\begin{table}[ht]
\caption{Values of the even-$J$ two-body matrix elements (m.e.) for the 
different interactions mentioned throughout the paper: quadrupole--quadrupole 
($Q\cdot Q$), surface delta (SDI), m.e. taken from experimental spectrum of 
$^{98}$Cd [V($^{98}$Cd)], and m.e. taken from experimental spectrum of 
$^{82}$Zr [V($^{82}$Zr)].} \label{tab:3}
\begin{tabular*}{.7\textwidth}[t]{@{\extracolsep{\fill}}ccccc}
\toprule
\multicolumn{1}{c}{$J$} & \multicolumn{1}{c}{$Q\cdot Q$} & \multicolumn{1}{c}
{SDI} & \multicolumn{1}{c}{V($^{98}$Cd)} & \multicolumn{1}{c}{V($^{82}$Zr)}\\
\colrule
0 & 0.0000 & 0.0000 & 0.0000 & 0.0000 \\
2 & 0.3485 & 2.0063 & 1.3947 & 0.4070 \\
4 & 0.9848 & 2.3149 & 2.0823 & 1.0408 \\
6 & 1.4848 & 2.4507 & 2.2802 & 1.8879 \\
8 & 1.1818 & 2.5415 & 2.4275 & 2.9086 \\
\botrule
\end{tabular*}
\end{table}

It should be said at the outset, however, that $Z=40,N=40$ is not a good closed
shell. Experiments by Lister et al.~\cite{letal87} showed that $^{80}$Zr is
strongly deformed. Skyrme--Hartree-Fock calculations by Bonche et 
al.~\cite{bfhkw85} and by Zheng and Zamick~\cite{zz91} are in agreement with
experiment. In Ref.~\cite{zz91} it was noted that, in the intrinsic deformed 
ground state of $^{80}$Zr, there were 12 nucleons in the $g_{9/2}$ shell. In
the spherical limit, there would not be any. For this reason, we will not show
single-$j$-shell fits to the Zr isotopes.

\section{Ground state spins}

For identical particles in the $g_{9/2}$ shell, the delta interaction yields a
ground state spin $I=j=9/2^+$ for odd--even or even--odd nuclei. In contrast, 
the $Q\cdot Q$ interaction yields $I=7/2^+$. Experimentally, it turns
out (as will be shown in the next section) that some nuclei have ground-state
spins $I=9/2^+$ and others have $I=7/2^+$. The latter nuclei are closer to the
$Z=40,N=40$ `closed shell', while the former are closer to the $Z=50,N=50$ 
closed shell. This shows that both the delta and $Q\cdot Q$ interactions are 
important for a proper description of these nuclei.

\section{The three and five particle (hole) spectra}

The nuclei we consider are broken into two groups: one in which $g_{9/2}$ 
protons are removed from a $Z=50,N=50$ core and a second in which $g_{9/2}$ 
neutrons are added to a $Z=40,N=40$ core. We shall see a significant and 
systematic difference in the behaviour in the two cases. The nuclei we consider
and the number of proton holes or neutron particles are shown in 
Table~\ref{tab:1}.

\begin{table}[ht]
\caption{Nuclei we consider in this work.} \label{tab:1}
\begin{tabular*}{.6\textwidth}[t]{@{\extracolsep{\fill}}ccccc}
\toprule
\multicolumn{2}{c}{$(Z=50,N=50)$ core} & & \multicolumn{2}{c}{$(Z=40,N=40)$ 
core} \\ \cline{1-2} \cline{3-5}
\# of holes & Nucleus & & \# of particles & Nucleus \\
2 & $^{98}$Cd & & 2 & $^{82}$Zr \\
3 & $^{97}$Ag & & 3 & $^{83}$Zr \\
5 & $^{95}$Rh & & 5 & $^{85}$Zr \\
7 & $^{93}$Tc & & 7 & $^{87}$Zr \\
\botrule
\end{tabular*}
\end{table}

\begin{figure}[ht]
\includegraphics[clip,scale=.5]{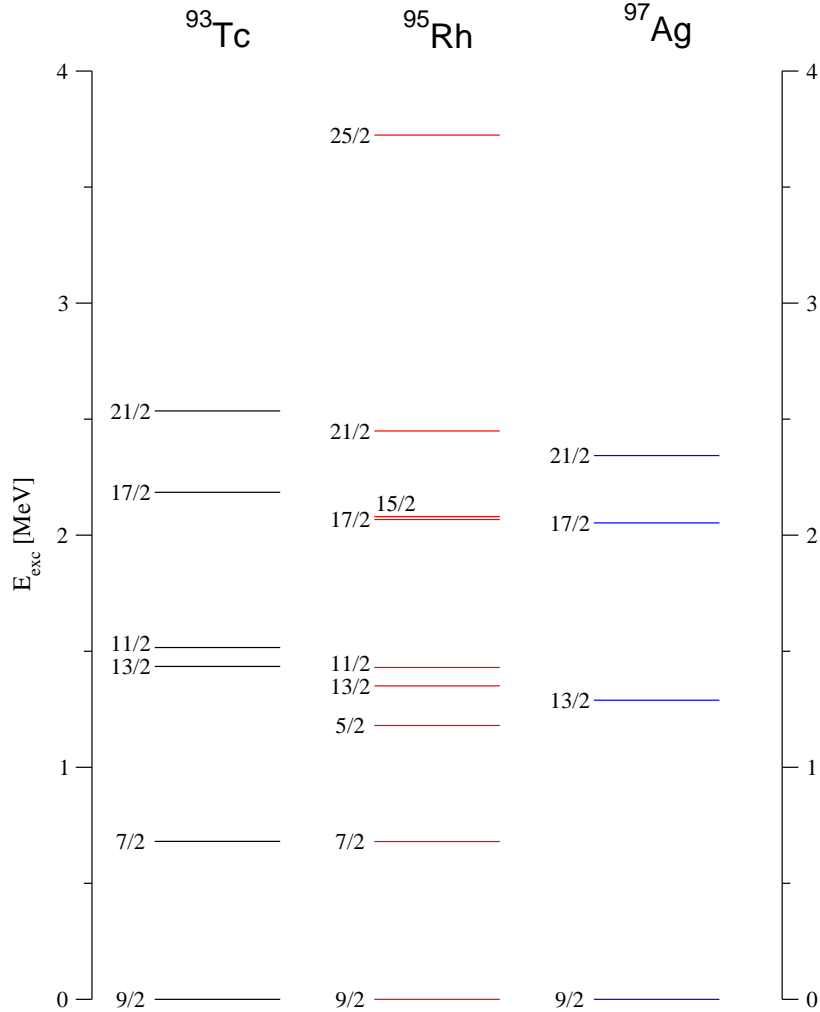}
\caption{Experimental energies of $^{93}$Tc, $^{95}$Rh, and $^{97}$Ag.}
\label{fig:50core-exp}
\end{figure}

\begin{figure}[ht]
\includegraphics[scale=.5,clip]{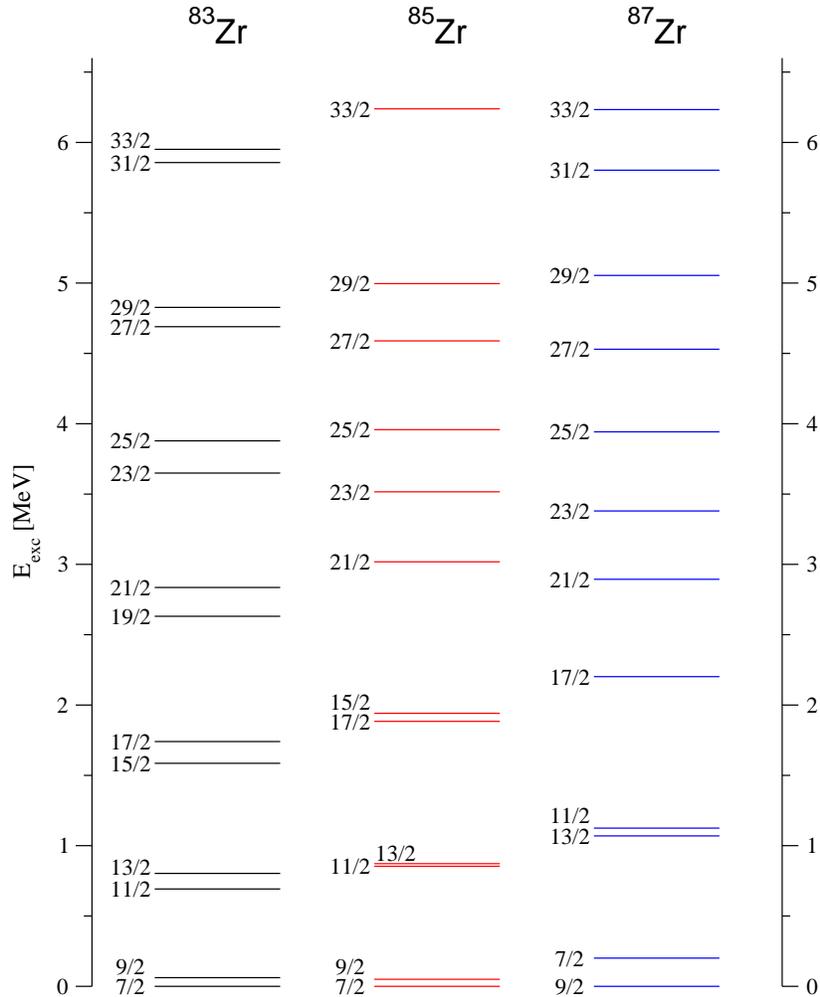}
\caption{Experimental energies of $^{83}$Zr, $^{85}$Zr, and $^{87}$Zr.}
\label{fig:40core-exp}
\end{figure}

In Fig.~\ref{fig:50core-exp} we show the empirical spectra of nuclei for 
$n=3,5$, and 7
protons removed from the $Z=50,N=50$ core---these are $^{97}$Ag, $^{95}$Rh, and
$^{93}$Tc, respectively. In Fig.~\ref{fig:40core-exp} we show the corresponding
spectra for $n=3,5$, and 7 neutrons relative to a $Z=40,N=40$ core (which we
had pointed out was deformed).

In an idealized world where the shell model worked perfectly, we would expect
the spectra of the three-particle system to be identical to that of the 
seven-particle system (i.e., three holes). Hence, $^{97}$Ag and $^{93}$Tc would
have the same spectrum. 
If furthermore the
interaction conserved seniority, then the spectrum of states with $v=1$ and 
$v=3$ would be the same for three particles and five particles.

We could go even further and say that, if the interaction for two proton holes
were the same as that for two neutrons, then the spectra of $^{93}$Tc, 
$^{97}$Ag, $^{83}$Zr, and $^{87}$Zr would all be the same. (If the spectra of
$^{93}$Tc and $^{83}$Zr are different, it does not mean that we have a 
violation of charge symmetry, of course). But that is really pushing the 
envelope.

Looking at the experimental spectra of Fig.~\ref{fig:50core-exp}, we can see 
that, although the agreement 
for three holes ($^{97}$Ag) and seven holes ($^{93}$Tc) is not perfect, it is 
still quite good. Also the fact that there is close agreement with $^{95}$Rh 
indicates that we are not far away from the seniority-conserving limit.


Looking at Fig.~\ref{fig:40core-exp}, we see that the spectra of the Zr 
isotopes are significantly different from 
those of $^{93}$Tc, $^{95}$Rh, and $^{97}$Ag. This undoubtedly is due to the 
fact that $Z=40,N=40$ is deformed. The $J=7/2^+$ ground state spins of 
$^{83}$Zr and $^{85}$Zr agree with the predictions of the $Q\cdot Q$ 
interaction, but not the delta interaction. 

Note the nearly degenerate doublet structure in the experimental spectrum of
$^{83}$Zr in Fig.~\ref{fig:40core-exp}, taken from the work of H\"uttmeier 
et~al.~\cite{hetal88}. The known doublets have angular momenta
$(7/2,9/2)$, $(11/2,13/2)$, $(15/2,17/2),\cdots$, up to $(47/2,49/2)$, although
we only show up to $(31/2,33/2)$. However, for a $j^3$ configuration of 
identical particles, there are no states with $J=J_\text{max}-1$ or $J>
J_\text{max}$, where $J_\text{max}=21/2$ for $j=9/2$. Hence, those states must 
have different configurations.

In the H\"uttmeier reference~\cite{hetal88}, a theoretical analysis using a 
Wood-Saxon cranking model was performed. A triaxial shape was predicted, which 
was the main cause of a signature splitting 
that leads to the deviation from a simple rotational spectrum. Discussions of 
signature splitting for triaxial nuclei can be found in several places, e.g., 
B.R.~Mottelson~\cite{m83}, Y.S.~Chen et al.~\cite{cfl83}, and I.~Hamamoto and
B.R.~Mottelson~\cite{hm8386}.

\section{Calculations with matrix elements from experiment}

We can get two-body matrix elements for a $Z=50,N=50$ core from the 
2-proton-hole spectrum of $^{98}$Cd---we will call it V($^{98}$Cd). If the
excitation energy of the lowest state of angular momentum $J$ ($J=0,2,4,6$,
and 8) is $E(J)$, then we make the association $\langle (\bar{g}_{9/2}^2)^J |V|
(\bar{g}_{9/2}^2)^J \rangle = E(J)$. And now we can proceed to do calculations 
for $n$ holes, with $n>2$. Note that, except for an overall constant, the 
hole--hole spectrum is the same as the particle--particle spectrum.

\begin{figure}[ht]
\includegraphics[scale=.5,clip]{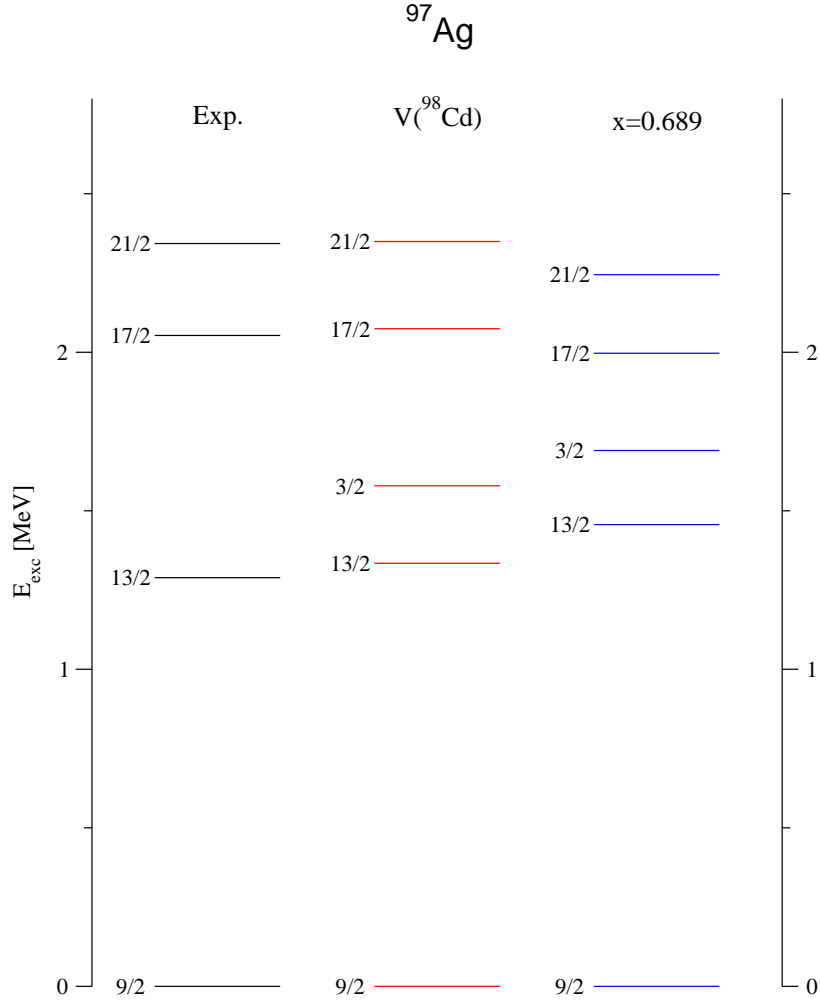}
\caption{Experimental and calculated spectra of $^{97}$Ag.}
\label{fig:97ag1}
\end{figure}

\begin{figure}[ht]
\includegraphics[scale=.5,clip]{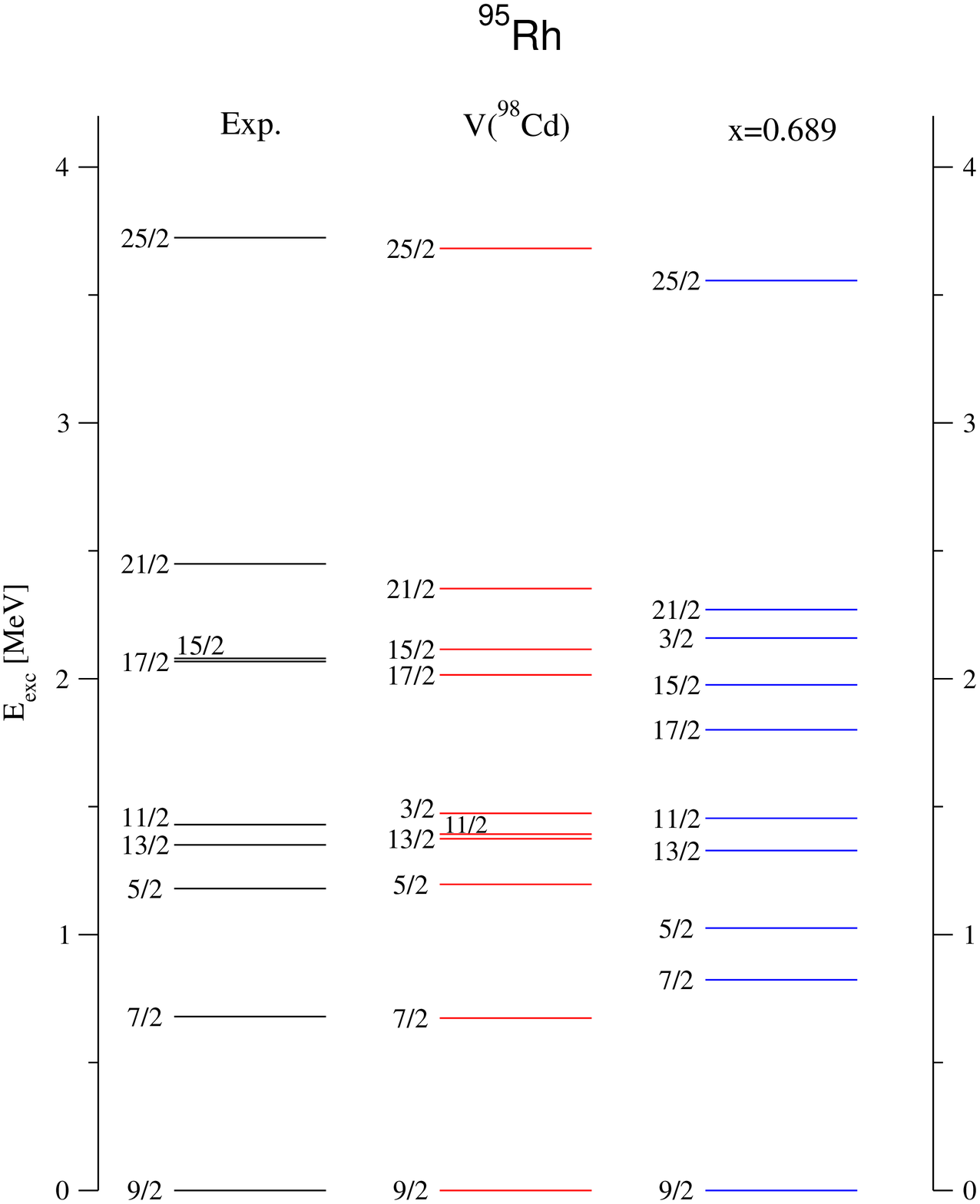}
\caption{Experimental and calculated spectra of $^{95}$Rh.}
\label{fig:95rh1}
\end{figure}

\begin{figure}[ht]
\includegraphics[scale=.5,clip]{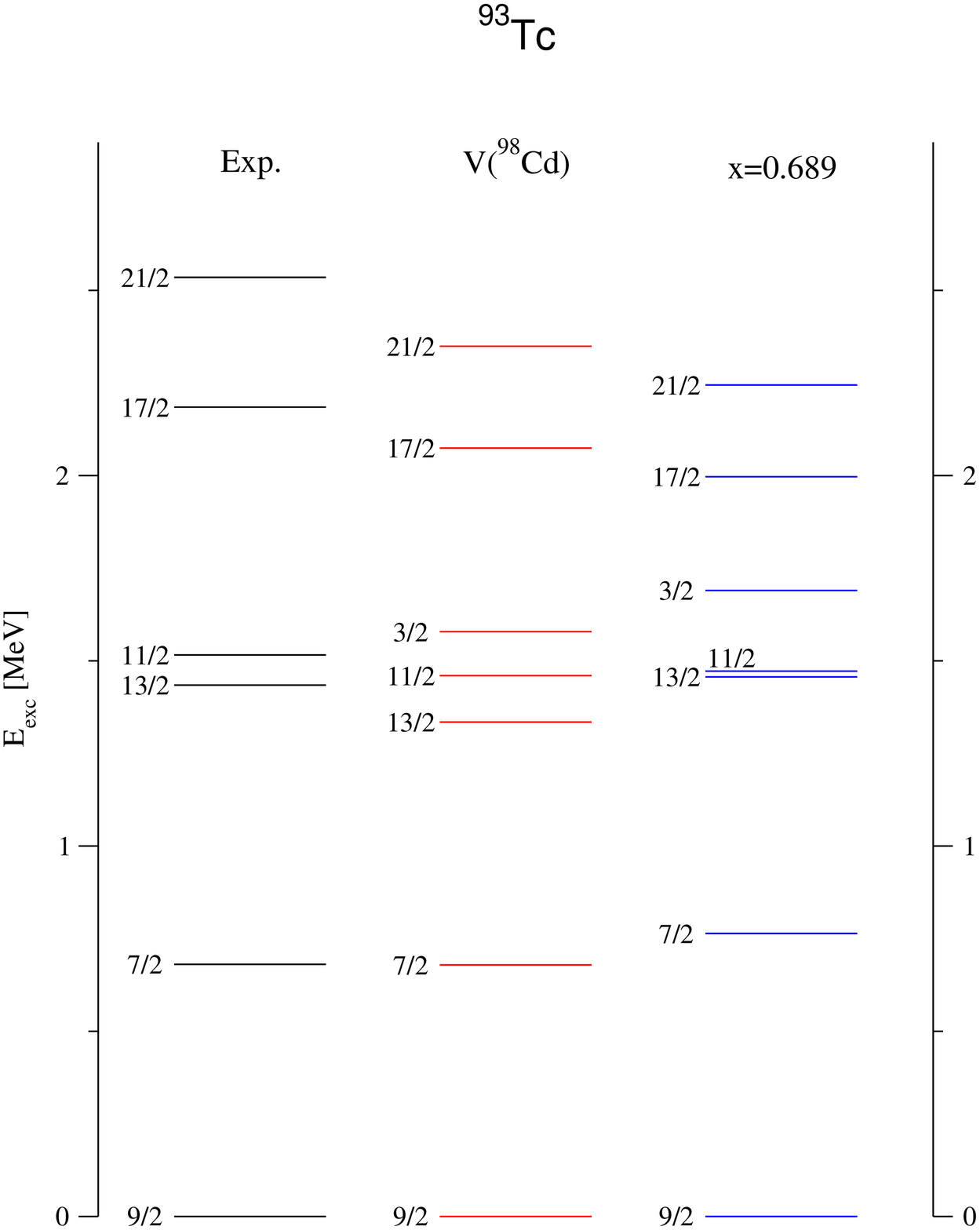}
\caption{Experimental and calculated spectra of $^{93}$Tc.}
\label{fig:93tc1}
\end{figure}

In Figs.~\ref{fig:97ag1}, \ref{fig:95rh1}, and \ref{fig:93tc1}, we show a 
comparison of the calculated spectra for 
V($^{98}$Cd) with experiment for $n=3,5$, and 7 proton holes corresponding to
$^{97}$Ag, $^{95}$Rh, and $^{93}$Tc, respectively. The results, although not
perfect, are quite reasonable considering the simplicity of the model.

\section{Linear combination of a Delta and $Q\cdot Q$ interaction}

The formula for the two-body matrix elements of the $Q\cdot Q$ interaction is
\begin{equation}
\langle [j j]^J |V_{Q\cdot Q}| [j j]^J \rangle = (-1)^J V_0 \frac{5}{4\pi} 
\langle r^2 \rangle_1 \langle r^2 \rangle_2 (2j+1) (j 2 \frac{1}{2} 0 | j 
\frac{1}{2})^2 
\begin{Bmatrix}
j & j & J \\ j & j & 2
\end{Bmatrix} ~.
\end{equation}

The formula for the surface delta interaction of Moszkowski and 
collaborators~\cite{gm65,am66,pam66} is
\begin{eqnarray}
\langle [j j]^{J T} |V_\text{SD}| [j j]^{J T} \rangle & = & 
- W_0 \frac{(2j+1)^2}{4(2J+1)} \\
 & & \times \left[ \{1+(-1)^T\} (j j \frac{1}{2} \frac{1}{2} | J 1)^2 +
\{1-(-1)^{J+T} \} (j j \frac{1}{2} -\frac{1}{2} | J 0)^2 \right] . \nonumber
\end{eqnarray}
In a single $j$ shell, there is no distinction between a delta interaction and 
a surface delta interaction.

\begin{figure}[ht]
\includegraphics[scale=.55,clip]{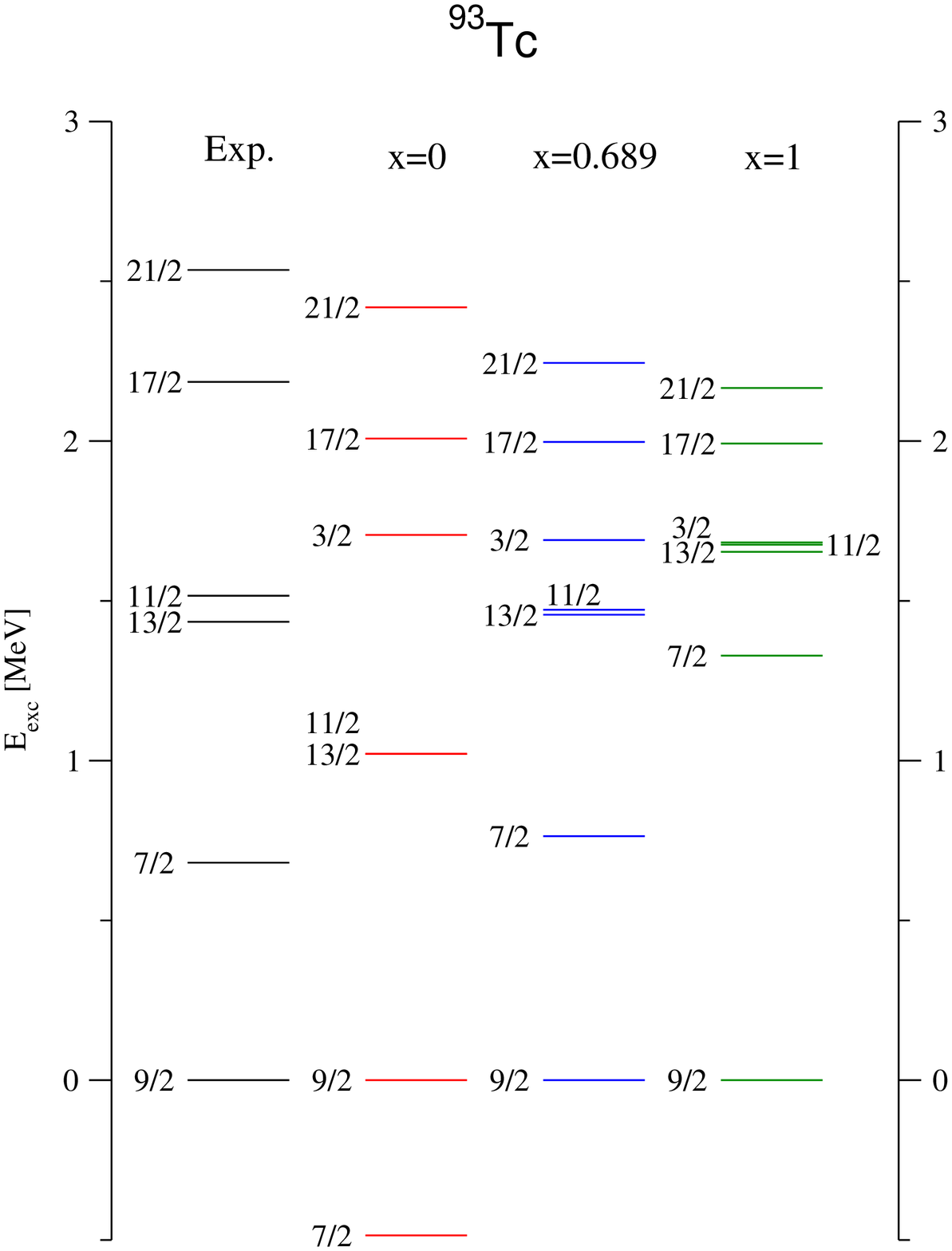}
\caption{Experimental and calculated spectra of $^{93}$Tc. $x=0$ means pure
$Q\cdot Q$ interaction; $x=1$, pure delta interaction; and $x=0.111$ is our
best linear combination of both interactions for this nucleus.}
\label{fig:93tc}
\end{figure}

\begin{figure}[ht]
\includegraphics[scale=.55,clip]{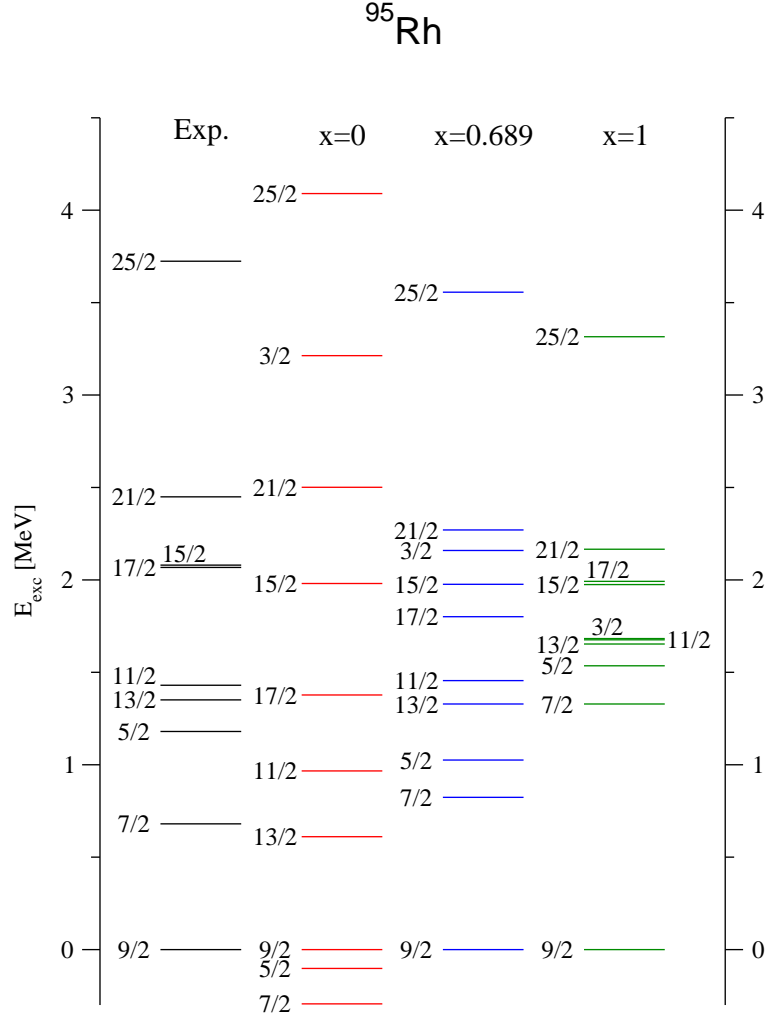}
\caption{Same as Fig.~\ref{fig:93tc} for $^{95}$Rh.}
\label{fig:95rh}
\end{figure}

\begin{figure}[ht]
\includegraphics[scale=.55,clip]{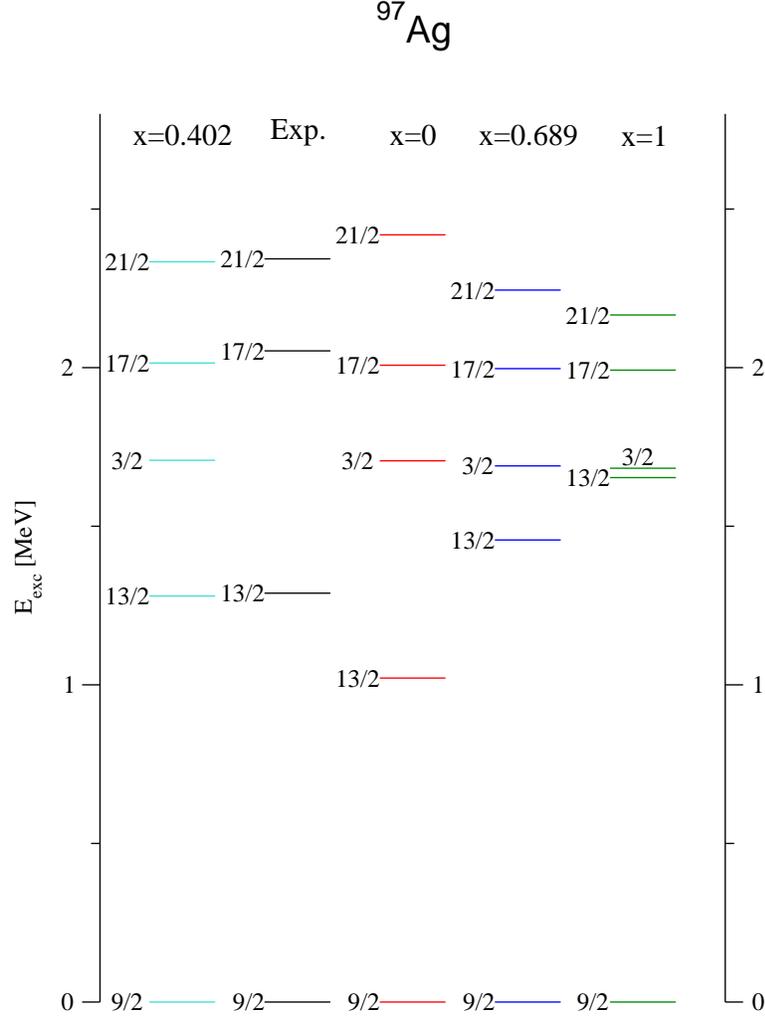}
\caption{Same as Fig.~\ref{fig:93tc} for $^{97}$Ag.}
\label{fig:97ag}
\end{figure}

In Figs.~\ref{fig:93tc}--\ref{fig:97ag}, we
give results for the $Q\cdot Q$ and delta interactions, choosing optimum
$V_0$ and $W_0$, respectively. Then, we form the linear combination $[x 
V_{Q\cdot Q} + (1-x) V_\text{SD}]$ and show the optimum $x$ to fit 
experiment. Thus, $x=0$ in the figures corresponds to pure $Q\cdot Q$, while
$x=1$ means pure delta. The values of $V_0$, $W_0$, and $x$ are shown in 
Table~\ref{tab:2}.

\begin{table}[ht]
\caption{Values of the optimum $V_0$, $W_0$, and $x$ (see text) for each 
isotope considered in the paper; $F(x)$ gives an estimate of how well our 
calculated energies fit experiment (the closer to zero, the better); G.s. 
stands for the experimental ground state.} \label{tab:2}
\begin{tabular*}{.8\textwidth}[t]{@{\extracolsep{\fill}}ccccccc}
\toprule
\multicolumn{1}{c}{$Z$} & \multicolumn{1}{c}{$N$} & \multicolumn{1}{c}{$x$}
 & \multicolumn{1}{c}{$F(x)$} & \multicolumn{1}{c}{$W_0$} & 
 \multicolumn{1}{c}{$V_0$} & \multicolumn{1}{c}{G.s.} \\
\colrule
43 & 50 & 0.6286 & 0.0830 & 0.4969 & 0.0296 & $9/2^+$ \\
45 & 50 & 0.6887 & 0.1865 & 0.5066 & 0.0270 & $9/2^+$ \\
47 & 50 & 0.4021 & 0.0016 & 0.5023 & 0.0272 & $9/2^+$ \\
40 & 43 & 0.0471 & 0.0307 & 0.3037 & 0.0190 & ($7/2^+$) \\
40 & 45 & 0.0495 & 0.4521 & 0.9060 & 0.0271 & $7/2^+$ \\
40 & 47 & 0.2808 & 0.2774 & 0.4624 & 0.0316 & $9/2^+$ \\
\botrule
\end{tabular*}
\end{table}

Because $^{80}$Zr is deformed, we do not show figures for single $j$ shell fits
to $^{83,85,87}$Zr. However, from Figs.~\ref{fig:93tc}--\ref{fig:97ag} (proton 
holes relative to $Z=50,N=50$), we can get some feeling for what is happening. 
In Fig.~\ref{fig:93tc} ($^{93}$Tc) we focus on the pure $Q\cdot Q$ ($x=0$) and 
surface delta ($x=1$) limits.

Whereas in $^{93}$Tc the $J=9/2^+$ is the lowest state, in $^{83}$Zr the 
$J=7/2^+$ is the lowest. The $Q\cdot Q$ interaction displays this 
feature---$E(7/2)<E(9/2)$---, but the surface delta does not. So, if we were
naively to try to fit the $^{83}$Zr spectrum with a $g_{9/2}^3$ configuration, 
we would need much more $Q\cdot Q$ than we needed for the fit to $^{93}$Tc. On
the other hand, some of the near doublet structure seen in the experimental 
spectrum of
$^{83}$Zr $(11/2,13/2)$ and $(15/2,17/2)$ is also a property of the delta 
interaction. However, introducing a lot of $Q\cdot Q$ will destroy the near 
degeneracy of these doublets. Thus, it is essentially impossible to fit both 
features of the $^{83}$Zr spectrum---a $J=7/2^+$ ground state and nearly 
degenerate doublets---with a combination of $Q\cdot Q$ and delta in a 
$g_{9/2}^3$ configuration.

\section{The $E(I_\text{max})-E(I_\text{min})$ splitting for $n=3$ and $n=5$ : 
$^{97}$A\lowercase{g} versus $^{95}$R\lowercase{h} and $^{83}$Z\lowercase{r} 
versus $^{85}$Z\lowercase{r}}

As mentioned in a previous section, the splitting $\Delta E = E(I_\text{max}=
21/2^+)-E(I_\text{min}=3/2^+)$ is the same for three particles as it is for 
five particles (or three holes and five holes) if one has a 
seniority-conserving interaction. However, for a pure $Q\cdot Q$ interaction, 
we have $\Delta E(n=5)=-\Delta E(n=3)$. 

Using the V($^{98}$Cd) interaction, we find
\begin{eqnarray}
\Delta E (n=3) & = & 0.77058 ~\text{ MeV} \nonumber ~, \\
\Delta E (n=5) & = & 0.87818 ~\text{ MeV} ~. \nonumber 
\end{eqnarray}
They are both positive, an indication that the seniority-conserving delta 
interaction is much more important than the seniority-violating $Q\cdot Q$ 
interaction.

I. Talmi had previously concluded, from an analysis of $h_{11/2}$ nuclei with
a closed shell of neutrons ($N=82$), that seniority conservation held to a high
degree~\cite{t93,t05}.

Unfortunately, for the $g_{9/2}$ nuclei that we are here considering 
($^{93}$Tc, $^{95}$Rh, $^{97}$Ag, as well as the zirconium isotopes $^{83}$Zr, 
$^{85}$Zr, and $^{87}$Zr), although the high spin states including $I=
21/2^+$ have been identified, the $I=3/2^+$ states have not been found yet. So
our analysis provides very strong motivation for an experimental search for the
$I=3/2^+$ states in $^{97}$Ag, $^{95}$Rh, and $^{93}$Tc, as well as for the Zr
isotopes.

For $^{83}$Zr and $^{85}$Zr, with a fitted interaction (despite misgivings of
using a single $j$ model space), we find for $\Delta E=E(I_\text{max})-
E(I_\text{min})$
\begin{eqnarray}
\Delta E(^{83}\text{Zr}) & = & 0.48742 ~\text{ MeV} ~, \nonumber \\
\Delta E(^{85}\text{Zr}) & = & -0.59355 ~\text{ MeV} ~. \nonumber
\end{eqnarray}
They have opposite signs, which shows that for these fitted interactions the
$Q\cdot Q$ interaction is much more important for this case---neutrons beyond
a $Z=40,N=40$ core---than it is for the case of proton holes relative to a
$Z=50,N=50$ core.

But it should be emphasized that the $I=3/2^+$ state is not part of the fit 
because it has not been identified experimentally. If more levels were known in
the Zr isotopes, and in particular the low spin level $I=3/2^+$ (but also 
$5/2^+$ and $1/2^+$), then the picture might change. We strongly urge that 
experimental work be done on all the nuclei considered here in order to locate 
the missing states, especially $I=3/2^+_1$ and also $5/2^+_1$.


\begin{acknowledgments}
We thank Igal Talmi for a critical and insightful reading of the first draft of
this work. Further comments from Naftali Auerbach are also appreciated, as well
as a program and input from Ben Bayman. We would like to acknowledge support 
from the Secretar\'{\i}a de Estado de Educaci\'on y Universidades (Spain) and 
the European Social Fund.
\end{acknowledgments}

\end{document}